\documentclass[12pt] {article}
\usepackage{latexsym}{\rm }
\usepackage{graphics}
\usepackage{graphpap}

\newcommand{\beq}{\begin{equation}}
\newcommand{\feq}[1]{\label{#1} \end{equation}}
\newcommand{\beqr}{\begin{eqnarray}}
\newcommand{\feqr}{\end{eqnarray}}
\def\non{\nonumber}
\def\noi{\noindent}

\newcommand{\rf}[1]{(\ref{#1})}

\def\np#1#2#3{Nucl. Phys. {\bf{B#1}} (#2) #3}

\def\prev#1#2#3{Phys. Rev. {\bf{D#1}} (#2) #3}
\def\prl#1#2#3{Phys. Rev. Lett. {\bf{#1}} (#2) #3}
\def\cqg#1#2#3{Class. Quant. Grav. {\bf{#1}} (#2) #3}

\def\plb#1#2#3{Phys. Lett. {\bf{B#1}} (#2) #3}
\def\mpl#1#2#3{Mod. Phys. Lett. {\bf{A#1}} (#2) #3}

\def\jhp#1#2#3{JHEP {\bf{#1}} (#2) #3}

\def\lmp#1#2#3{Lett. Math. Phys. {\bf{#1}} (#2) #3}

\renewcommand{\thefootnote}{\fnsymbol{footnote}}

\begin{document}

\begin{center}


{\Large \bf Closed Bosonic String Partition Function in Time Independent Exact PP-Wave Background}\\
[4mm]

\large{Agapitos Hatzinikitas} \\ [5mm]

{\small University of Crete, \\
Department of Applied Mathematics, \\
L. Knosou-Ambelokipi, 71409 Iraklio Crete,\\
Greece, \\
Email: ahatzini@tem.uoc.gr}\\ [5mm]

\large{and} \\ [5mm]

\large{Ioannis Smyrnakis} \\ [5mm]

{\small University of Crete, \\
Department of Applied Mathematics, \\
L. Knosou-Ambelokipi, 71409 Iraklio Crete,\\
Greece, \\
Email: smyrnaki@tem.uoc.gr} \vspace{5mm}
\end{center}

\begin{abstract}
  The modular invariance of the one-loop partition function of the closed bosonic string in 
four dimensions in the
presence of certain homogeneous exact pp-wave backgrounds is
studied.  In the absence of an axion field the partition function is
found to be modular invariant. In the presence
of an axion field modular invariace is broken. This can be attributed
to the light-cone gauge which breaks the symmetry in the $\sigma$-, $t$-directions. 
Recovery of this broken modular invariance suggests the introduction of twists 
in the world-sheet directions. However, one needs to go beyond the light-cone gauge to 
introduce such twists. 
\end{abstract}
\newpage

\section{Introduction}
\renewcommand{\thefootnote}{\arabic{footnote}}
\setcounter{footnote}{0}
\par
It has been known for some time \cite{peres} that there are certain
solutions to the vacuum Einstein equations in four dimensions, with
a covariantly constant null Killing vector, that can be interpreted
as plane fronted gravitational waves, the so called pp-waves. These
are in Brinkmann coordinates and in four dimensions of the form
\beqr \label{1} ds^2=-dX^+dX^-+ F(X^+,X_i){(dX^+)}^2+
\sum_{i=1}^{2}(dX_i)^2 \feqr where $\sum_{i=1}^{2}(dX_i)^2$ denotes
the standard metric in the Euclidean space $E^2$ and
 the metric is a solution to the vacuum Einstein field equations if and only if
\beqr \label{1a}
\partial_T^2F(X^+,X_i)=0.
\feqr \noi A particular solution of \rf{1a} is
$F(X^+,X_i)=C_{ij}\left(X^+\right)X^i X^j$ with $Tr
\left(C_{ij}\right)=0$.
\par It was shown in \cite{amati} that pp-waves are exact
solutions of the spacetime string equations to all orders of
perturbation theory.  So if this metric is used as a spacetime
metric in string theory, the $\beta$-functions are zero, so
conformal invariance is preserved \cite{callan}. This class of
solutions can be extended by the introduction of an antisymmetric
tensor and a dilaton field so that the $\beta$-functions remain zero
\cite{horowitz}. The axion field strength corresponding to the
antisymmetric tensor is given by \beqr \label{2} H_{\mu \nu
\rho}=A_{ij}(V+Z)l_{[ \mu}\nabla_{\nu}X_i \nabla_{\rho ]}X_j \feqr
\noi where $l_{\mu}=\partial_{\mu}X^+$ is the null Killing vector in
the coordinates $V=\frac{1}{2}(X^++X^-), Z=\frac{1}{2}(X^+-X^-),
X^i, \: i=1,2$. The condition of Weyl invariance in two dimensions
now becomes \beqr \label{3} R_{\mu\nu}-\frac{1}{4}H_{\mu \rho
\sigma}H_{\nu}^{\; \; \rho \sigma} -\nabla_{\mu}\nabla_{\nu}\Phi=0.
\feqr \noi Equation \rf{3} is rewritten as \beqr \label{3a}
-\frac{1}{2}l_{\mu}l_{\nu}\partial_i\partial^iF(V+Z, X^i)
-\frac{1}{36}l_{\mu}l_{\nu}A_{ij}(V+Z)A^{ij}(V+Z)
-\partial_{\mu}\partial_{\nu}\Phi(V+Z)=0 \feqr \noi where only the
$\mu=\nu=(V,Z)$ components contribute. They both lead to the
equation \beqr \label{3b}
\partial_T^2F + \frac{1}{18}A_{ij}A^{ij}+ 2\Phi^{''}=0
\feqr \noi where $\Phi^{''}$ refers to differentiation of the
dilaton field w.r.t. $X^+$. The action corresponding to the above
exact solution of the spacetime string equations is \beqr
S=\frac{1}{2\pi}\int d\sigma dt \left(h^{ab} \partial_a X^+
\partial_b X^- - h^{ab} \partial_a X_i \partial_b X_i -h^{ab}F
\partial_a X^+ \partial_b X^+ -\frac{1}{3} A_{ij} X_j \partial_a X^+
\partial_b X_i \epsilon^{ab} \right) \label{4} \feqr \noi where
$\alpha'=1/2$, $h^{ab}=\sqrt{-g}g^{ab}$ and $\epsilon^{t
\sigma}=-1$. Note that $h=deth_{ab}=-1$. This action as it stands is
not quadratic in the string coordinates so it is difficult to
manipulate. For a particular class of pp-waves, the exact plane
waves, it turns out that by choosing the light-cone gauge it is
possible to make this action quadratic and hence quantize it. For
these exact plane waves $F(X^+,X^i)=C_{ij}(X^+)X^i X^j$.
Interestingly enough the transverse string coordinates become
massive bosonic fields. Nevertheless the conformal invariance is
manifested in the partition function. The modular invariance of the
partition function for the Nappi-Witten model \cite{nappi} has been
investigated in \cite{tadashi}.
\par Of particular interest are two families of exact plane waves that possess an extra Killing
vector capable of generating translations in $X^+-$direction. These
are called homogeneous exact plane waves. Their metrics in Brinkmann
coordinates are given by \cite{blau} \beqr
ds^2=-dX^+dX^-+\left(e^{X^+ f}A_0 e^{-X^+
f}\right)_{ij}w_iw_j(dX^+)^2 + (d\vec{w})^2 \label{4a} \feqr \beqr
ds^2=-dX^+dX^-+\left(e^{f \ln X^+ }A_0 e^{-f \ln X^+
}\right)_{ij}w_iw_j \frac{(dX^+)^2}{(X^+)^2} + (d\vec{w})^2
\label{4b} \feqr \noi where $A_0$ is a constant symmetric matrix and
$f$ is a constant antisymmetric matrix. The first family consists of
geodesically complete metrics having no singularities. The second
family has a null singularity at $X^+=0$.
\par The second family of homogeneous exact plane waves has been investigated in
 \cite{Russo}. It has been shown that the closed string theory based on this special metric
is an exactly solvable model meaning that it is possible to find
explicitly the solution to the classical string equations, perform a
canonical quantization, determine the spectrum of the Hamiltonian
operator and compute some simple scattering amplitudes.
\par Here we will be concerned mostly with the first family although the covariant form of the
action given at the end of Section 4 holds for any plane fronted
wave. If we go to a rotating frame through the transformation \beqr
w_i=\left(e^{X^+ f}\right)^i_k X^k \label{4c} \feqr \noi the metric
then becomes \beqr
ds^2==-dX^+dX^-+(A_0-f^2)_{ij}X^iX^j(dX^+)^2+(dX^i)^2-2f_{ij}X^i
dX^j dX^+. \label{4d} \feqr \noi This is a more convenient frame for
quantization since neither $A_0$ nor f depend on $X^+$. Early
studies of strings propagating on subclasses of these backgrounds
have been made by \cite{brooks, vega, tseytlin}.
\par Our aim in the present paper is to study the modular properties of the partition function
for the closed bosonic string in four dimensions which propagates on
the first family exact plane wave background with $f_{ij}=0$, in the
presence of the antisymmetric tensor and the dilaton field. To
achieve this we organise the paper as follows.
\par In \textit{Section 2} the canonical momenta are used to write the phase-space
action before fixing the light-cone gauge. From this we read off the
phase-space Hamiltonian in the light-cone gauge. Next we specialize
to time-independent exact plane waves. The transverse string
coordinates and the canonical momenta are decomposed into oscillator
modes. Finally each oscillator Hamiltonian is diagonalized through a
canonical transformation.
\par In \textit{Section 3} the partition function for $A_{ij}=0$ is computed in
the path integral formalism by expanding the transverse fields,
subject to free-theory boundary conditions, in modes and then
integrating them out. Following a deformed zeta function
regularization scheme one can prove that the partition function is
explicitly modular invariant, although it does not split into a
finite sum of holomorphic times antiholomorphic blocks in any
obvious way.
\par In \textit{Section 4} the now non zero axion 
field $A_{ij}$ is interpreted as an $O(2)$ worldsheet gauge field in the $\sigma$-direction. 
This is shown to break the modular
invariance due to inequivalence of the 
$\sigma$- and $t$-directions in the worldsheet,
for which the light-cone gauge is responsible. The necessity for
modular invariance seems to indicate
that we need the introduction of twists in both the 
$\sigma$- and $t$-directions. However, to justify
this, one needs to go beyond the light-cone gauge.
\section{Quantization}

\par The canonical momenta corresponding to the action \rf{4} are
\beqr
\mathcal{P}_+ &=& \frac{\partial L}{\partial \dot{X}^-}= \frac{1}{2\pi} h^{a0}\partial_a X^+ \non \\
\mathcal{P}_- &=& \frac{\partial L}{\partial \dot{X}^+}=
\frac{1}{2\pi}
\left(h^{a0} \partial_a X^- -2F\mathcal{P}^+ + \frac{1}{3} A_{ij}X_j X_i^{'}\right) \non \\
\mathcal{P}_i &=& \frac{\partial L}{\partial \dot{X}_i}= -
\frac{1}{2\pi} \left(2 h^{a0}\partial_a X_i + \frac{1}{3} A_{ij} X_j
X^{+ '} \right). \label{5} \feqr \noi The canonical momenta \rf{5}
conjugate to the bosonic coordinates is the generalization of those
used in \cite{metsaev} when an axion field is switched on.
Substituting in the action we get \beqr S &=& \frac{1}{2\pi}\int
d\sigma dt \Biggl[ 2\pi \mathcal{P}_+ \dot{X}^- + 2\pi \mathcal{P}_-
\dot{X}^+ + 2\pi \mathcal{P}_i \dot{X}_i +
\frac{h^{01}}{h^{00}}\left(2\pi \mathcal{P}_+ X^{- '} + 2\pi
\mathcal{P}_- X^{+ '}
+ 2\pi \mathcal{P}_i X_i^{'} \right) \non \\
&-& \frac{1}{h^{00}}\left(4\pi^2 \mathcal{P}_+ \mathcal{P}_- +
4\pi^2 F (\mathcal{P}_+)^2 - \pi^2 (\mathcal{P}_i)^2 +
X^{+ '}X^{- '} - F (X^{+ '})^2 - (X_i^{'})^2 \right) \non \\
&+& \frac{1}{h^{00}}\left(\frac{2\pi}{3}A_{ij}\mathcal{P}_+ X_j
X_i^{'} + \frac{2\pi}{6} A_{ij} \mathcal{P}_i X_j X^{+ '} +
\left(\frac{1}{6}A_{ij}X_j X^{+ '} \right)^2  \right) \Biggl].
\label{6} \feqr In the conformal gauge $h^{00}=-1$ and $h^{10}=0$ so
the action becomes \beqr S &=& \frac{1}{2\pi}\int d\sigma dt \Biggl[
2\pi \mathcal{P}_+ \dot{X}^-
+ 2\pi \mathcal{P}_- \dot{X}^+ + 2\pi \mathcal{P}_i \dot{X}_i \non \\
&+& \left(4\pi^2 \mathcal{P}_+ \mathcal{P}_- + 4\pi^2 F
(\mathcal{P}_+)^2 - \pi^2 (\mathcal{P}_i)^2 +
X^{+ '}X^{- '} - F (X^{+ '})^2 - (X_i^{'})^2 \right) \non \\
&-& \left(\frac{2\pi}{3}A_{ij}\mathcal{P}_+ X_j X_i^{'} +
\frac{2\pi}{6} A_{ij} \mathcal{P}_i X_j X^{+ '} +
\left(\frac{1}{6}A_{ij}X_j X^{+ '} \right)^2  \right) \Biggl].
\label{8} \feqr \noi From this we can read off the Hamiltonian in
the light-cone gauge. Setting $X^+=p_+ t$ we obtain \beqr \label{9}
H=\int \frac{d\sigma}{2\pi} \left[2\pi p_+ \mathcal{P}_- + \pi^2
(\mathcal{P}_i)^2 + {X_i}' {X_i}' -F p_+^{2}
-\frac{p_+}{3}A_{ij}{X_i}' X_j \right]. \feqr
\par Next we specialize to the time independent exact plane waves. For these the
polarisation tensor is given by $C_{ij}=\left(\matrix{W_1
+\tilde{\Phi}  & -W_2  \cr -W_2  & -W_1 +\tilde{\Phi} }\right)$ and
we have \beqr \label{10} F(X^+,X_i)=C_{ij}X^i X^j= W_{1}(X^2_1
-X^2_2)-2W_2 X_1 X_2 + \tilde{\Phi}(X_1^2 +X_2^2). \feqr \noi Note
that $\partial^2_T F=4\tilde{\Phi}$. Setting $A_{ij}=A\epsilon_{ij}$
the spacetime field equation \rf{3b} demands that \beqr \label{11}
\tilde{\Phi}=-\frac{1}{36} A^2 - \frac{1}{2}\Phi''. \feqr \noi
Substituting \rf{10} into \rf{9} we obtain \beqr
H &=& \frac{1}{2\pi}\int d\sigma \Biggl[ 2\pi p_+ \mathcal{P}_- + \pi^2 (\mathcal{P}_i)^2 + X_i^{2 '} \non \\
&-& p_+^2 \left( W_{1}(X^2_1 -X^2_2)-2W_2 X_1 X_2 -
\left(\frac{1}{36} A^2 + \frac{1}{2}\Phi'' \right)
(X_1^2 +X_2^2)\right) \non \\
&-& \frac{p_+}{3}A \left({X_1}' X_2 - {X_2}' X_1  \right) \Biggl].
\label{12} \feqr
\par We proceed by expanding $X^i$ and $\mathcal{P}^i$ in oscillator modes
\beqr
X^i(\sigma, t) &=& \frac{1}{\sqrt{2}}\sum_{n \in Z}X_n^i (t) e^{in\sigma} \non \\
\mathcal{P}^i (\sigma, t) &=& \frac{1}{\pi \sqrt{2}}\sum_{n \in
Z}P_n^i (t) e^{-in\sigma}. \label{13} \feqr \noi Reality of $X^i$,
$\mathcal{P}^i$ demands that $X^i_{-n}(t)=\bar{X}^i_n (t)$,
$P^{i}_{-n}(t)=\bar{P}^{i}_{n}(t)$. This implies in particular that
$X^i_0(t)$, $P^i_0(t)$ are real. The commutation relations of the
$\hat{X}^i_n, \hat{P}^j_m$ in the operator formalism are given by
\beqr [\hat{X}^i_n, \hat{P}^j_m]=i\delta^{ij}\delta_{nm}.
\label{13a} \feqr \noi The Hamiltonian now becomes \beqr H=2\pi
p_+p_- + H_0 + H_{osc.} \label{14} \feqr \noi where $p_-$ is the
average value of the momentum density $\mathcal{P}_-$, \beqr
H_0=\frac{1}{2} (P^i_0)^2 - \frac{1}{2} p_+^2 W_{1}((X^1_0)^2 -
(X^2_0)^2) +p_+^2 W_2 X^1_0 X^2_0 + \frac{1}{2}p_+^2
\left(\frac{1}{36} A^2 + \frac{1}{2}\Phi''\right) (X^i_0)^2
\label{15} \feqr \noi and \beqr H_{osc.} &=& \frac{1}{2}\sum_{n \in
Z -\{0\}} \Biggl[ |P^i_n|^2 + n^2 |X^i_n|^2
- p_+^2 W_1 (|X^1_n|^2 - |X^2_n|^2)+ p_+^2 W_2 (X^1_n X^2_{-n}+ X^1_{-n} X^2_n) \non \\
&+& p_+^2 \left(\frac{1}{36} A^2 + \frac{1}{2}\Phi''\right)
|X^i_n|^2 - in \frac{p_+}{3}A (X^1_n X^2_{-n} - X^1_{-n}X^2_n)
\Biggl]. \label{16} \feqr \noi If we define as in \cite{horowitz}
\beqr
\phi_1 &=& p_+^2 \left(-W_1 + \frac{A^2}{36}+ \frac{\Phi''}{2} \right) \non \\
\phi_2 &=& p_+^2 \left(W_1 + \frac{A^2}{36}+ \frac{\Phi''}{2} \right) \non \\
\rho &=& W_2 p_+^2 \non \\
\lambda &=& \frac{1}{3} p_+ A \label{17} \feqr \noi we have \beqr
H_0 &=& \frac{1}{2} (P^i_0)^2 + \frac{1}{2} \left( X^1_0 \,X^2_0
\right) \left( \matrix{ \phi_1 & \rho \cr \rho & \phi_2 } \right)
\left( \begin{array}{c} X^1_0 \\
X^2_0 \end{array} \right) \non \\
H_{osc.} &=& \frac{1}{2} \sum_{n \in Z-\{0\}} \Biggl[ |P^i_n|^2 +n^2
|X^i_n|^2 + \left(X^{1 \dagger}_n \, X^{2 \dagger}_n \right) \left(
\matrix{ \phi_1 & (\rho + in\lambda) \cr (\rho -in\lambda) & \phi_2
} \right)
\left( \begin{array}{c} X^1_n \\
X^2_n \end{array} \right) \Biggl]. \label{18} \feqr \noi It is
possible to transform canonically the phase-space variables so as to
diagonalize the Hamiltonian. Let \beqr \tilde{X}^i_n =
M^{\dagger}_{ij}X^j_n \quad  \tilde{P}^i_n = M_{ij}P^j_n \label{19}
\feqr \noi where \beqr (M_{ij})=i \sqrt{\rho^2 + n^2 \lambda^2}
\left( \matrix{  \alpha_- e^{i\theta}& \alpha_+ \cr \alpha_+ &
-\alpha_- e^{-i\theta}} \right), \label{20} \feqr \noi \beqr
\alpha_{\pm} &=& \frac{1}{\sqrt{G_{\pm}^2 + \rho^2 + n^2 \lambda^2}} \non \\
G_{\pm} &=& \pm \left(\frac{\phi_1 -\phi_2}{2}\right) +
\sqrt{\left(\frac{\phi_1 -\phi_2}{2}\right)^2 + \rho^2 + n^2
\lambda^2 } \label{21} \feqr \noi and $\theta=arg(\rho -in\lambda)$.
The diagonalized Hamiltonians become \beqr H_0 &=& \frac{1}{2}
(\tilde{P}^i_0)^2 +
\frac{1}{2} \left( S_{0,+}^2 (\tilde{X}^1_0)^2 + S_{0,-}^2 (\tilde{X}^2_0)^2 \right)  \non \\
H_{osc.} &=& \frac{1}{2} \sum_{n \in Z- \{0\}} \Biggl[
|\tilde{P}^i_n|^2 + n^2 |\tilde{X}^i_n|^2 +
 S_{n,+}^2 |\tilde{X}^1_n|^2 + S_{n,-}^2 |\tilde{X}^2_n|^2 \Biggl] \non \\
&=& \sum_{\stackrel{n > 0}{i=1,2}} H^i_n \label{22} \feqr \noi where
\beqr H^i_n=\tilde{P}^i_n \tilde{P}^i_{-n} + (\omega^i_n)^2
\tilde{X}^i_n \tilde{X}^i_{-n}, \label{22a} \feqr \beqr S_{n,
\pm}^2= \left(\frac{\phi_1 +\phi_2}{2}\right) \pm
\sqrt{\left(\frac{\phi_1 - \phi_2}{2}\right)^2 + \rho^2 + n^2
\lambda^2 } \label{23} \feqr \noi and $(\omega^{1,2}_n )^2= n^2
+S^2_{n, \pm}$. Here we have assumed that the gravitational wave
amplitudes are small compared to $\Phi''$ so as to have positivity
of $S_{n, \pm}^2$. In case the gravitational waves have large
amplitudes then the average number of excitation modes of the string
diverges exponentially and a string singularity appears
\cite{horowitz}.


\section{Partition Function when $A=0$}
In the particular case of $A=0$ we have $\phi_1 = p_+^2 \left(-W_1 +
\frac{\Phi''}{2} \right)$, $\phi_2 = p_+^2 \left(W_1 +
\frac{\Phi''}{2} \right)$, $\rho = W_2 p_+^2$ and $\lambda = 0$, so
$S_{\pm}^2= \left(\frac{\phi_1 +\phi_2}{2}\right) \pm
\sqrt{\left(\frac{\phi_1 - \phi_2}{2}\right)^2 + \rho^2 }$ becomes
independent of $n$. Time independence of $\phi_1$ , $\phi_2$, $\rho$
demands that the dilaton is at most quadratic in $X^+$,
$\Phi(X^+)=c_1 (X^+)^2 + c_2 X^+ + c_3$ and that $W_1$, $W_2$ are
independent of $X^+$. So we have \beqr S_{\pm}^2= p^2_+ \left(c_1
\pm  \sqrt{W_1^2 + W_2^2 } \right)= p^2_+ \tilde{S}^2_{\pm}.
\label{23a} \feqr
\par The partition function now becomes
\beqr \mathcal{Z}_{A=0}=\int_{\mathcal{F}} \frac{d\tau
d\bar{\tau}}{\tau_2}Z_{A=0}(\tau, \bar{\tau}) \label{23b} \feqr \noi
where \beqr Z_{A=0}(\tau, \bar{\tau})= C \int dp_+ dp_-
Tr\left(e^{-2i\pi \tau_2 \hat{H}} e^{2i\pi \tau_1 \hat{\Pi}}
\right), \label{24} \feqr \noi $\hat{\Pi}=\sum_{i=1}^2
\sum_{n=0}^{\infty} \hat{\Pi}^i_n$ is the momentum operator of the
string, $\hat{H}=2\pi p_+p_- + \sum_{i=1}^2
\sum_{n=0}^{\infty}\hat{H}^i_n + \frac{1}{2}\sum_{n\in Z}\sqrt{n^2
+S^2_{+}}+ \frac{1}{2}\sum_{n\in Z}\sqrt{n^2 +S^2_{-}}$ is the
normal ordered Hamiltonian and $\mathcal{F}$ is the fundamental
domain of the modular transformations. Substituting the above
operators in \rf{24} and using the results of appendix A we get, in
Euclidean time, that \beqr Z_{A=0}(\tau, \bar{\tau})&=& C \int dp_+
dp_- e^{-2\pi \tau_2 p_+ p_-} \prod_{i=1}^2 \left( Z^i_0
\prod_{n>0}Z^i_n \right)
\non \\
&=& C \int dp_+ dp_- e^{-2\pi \tau_2 p_+ p_-} \prod_{i=1}^2 \left(
det^{-\frac{1}{2}}(D^i_0) \prod_{n>0} det^{-1}(D^i_n)
\right) \non \\
&=& C \int dp_+ dp_- e^{-2\pi \tau_2 p_+ p_-} \prod_{i=1}^2 \prod_{n
\in Z} det^{-\frac{1}{2}}(D^i_n) \label{25} \feqr \noi where $Z^i_n$
for $n\geq 0$ are given in appendix A. Using now the determinant
formulae derived in appendix B we have \beqr \prod_{n \in Z}det
(D^i_n) = \prod_{n \in Z} e^{2\pi \omega^i_n \tau_2} (1-e^{-2\pi
\omega^i_n \tau_2 + 2i\pi n \tau_1}) (1-e^{-2\pi \omega^i_n \tau_2 -
2i\pi n \tau_1}) = e^{4\pi \tau_2 \Delta_i} f_i^2 (\tau, \bar{\tau})
\label{26} \feqr \noi \beqr
\Delta_{(1,2)}=\Delta_{p_+\tilde{S}_{(+,-)}} = \frac{1}{2}\sum_{n
\in Z} \omega^{i}_n = \frac{1}{2} \sum_{n \in Z} \sqrt{n^2 + p_+^2
\tilde{S}_{(+,-)}^2}= -\frac{1}{2\pi^2} \sum_{n=1}^{\infty}
\int_{0}^{\infty} ds e^{-n^2 s - \frac{p^2_+ \tilde{S}^2_{(+, -)}
\pi^2}{s}} \label{27} \feqr \noi and \beqr f_i(\tau,
\bar{\tau})=\prod_{n \in Z} (1-e^{-2\pi \omega^i_n \tau_2 + 2i\pi n
\tau_1}). \label{28} \feqr \noi Note that in formula \rf{27} we have
regularized the sum by analytically continuing the formula (see
appendix C) \beqr \sum_{n=1}^{\infty} \frac{1}{(n^2 + c^2)^{\nu}}=
-\frac{1}{2c^{2\nu}} + \frac{\sqrt{\pi}}{2c^{2\nu -1}\Gamma(\nu)}
\left[ \Gamma(\nu -\frac{1}{2})+4\sum_{n=1}^{\infty} (\pi n c)^{\nu
- \frac{1}{2}} K_{\nu - 1/2}(2\pi nc)\right] \label{29} \feqr \noi
to $\nu=-1/2$ and dropping the infinite uniform vacuum energy
$\frac{\sqrt{\pi}}{2c^{2\nu -1}\Gamma(\nu)}\Gamma(\nu -\frac{1}{2})$
\cite{farina}. We also make use of the fact that \cite{ryzhik} \beqr
K_{-1}(z)=\frac{1}{z} \int_{0}^{\infty}e^{-t-\frac{z^2}{4t}}dt.
\label{30} \feqr \noi This regularized $\Delta_{i}$ is the Casimir
energy of the theory. The regularization procedure we followed
corresponds to a deformed zeta function regularization. Using the
notation of deformed modular forms as defined in appendix D we have
$\prod_{n \in Z}det (D^{(1,2)}_n) = \hat{\eta}^2_{p_+ \tilde{S}_{(+,
-)}}(\tau, \bar{\tau})$ so \beqr Z_{A=0}(\tau,\bar{\tau}) = C \int
dp_+ dp_- e^{-2\pi \tau_2 p_+ p_-} \left[ \hat{\eta}_{p_+
\tilde{S}_+}(\tau, \bar{\tau})
 \hat{\eta}_{p_+ \tilde{S}_-}(\tau, \bar{\tau}) \right]^{-1}.
\label{31} \feqr
\par The function $\hat{\eta}_{c}(\tau, \bar{\tau})$ has the modular properties \cite{green}
\beqr \label{32}
\hat{\eta}_{c}(\tau +1, \bar{\tau}+1) &=&  \hat{\eta}_{c}(\tau, \bar{\tau})  \non \\
\hat{\eta}_{c}(-\frac{1}{\tau}, -\frac{1}{\bar{\tau}}) &=&
\hat{\eta}_{c/|\tau|}(\tau, \bar{\tau}) \feqr \noi and in the limit
$c\rightarrow 0$ it degenerates according to \beqr \hat{\eta}_0^R
(\tau, \bar{\tau}) = \eta (\tau) \overline{\eta(\tau)}. \label{33}
\feqr \noi Here the symbol $R$ means that we have regularized
$\hat{\eta}_0 (\tau, \bar{\tau})$ by dropping a zero factor in the
limit $c\rightarrow 0$.
\par Because
of the modular properties \rf{32} and the fact we integrate over
$p_+$, $p_-$ we have that the partition function transforms
according to \beqr Z_{A=0}(\tau +1, \bar{\tau} +1)=Z_{A=0}(\tau ,
\bar{\tau}) \qquad Z_{A=0}(-\frac{1}{\tau}, -\frac{1}{\bar{\tau}})=
|\tau|^2 Z_{A=0}(\tau , \bar{\tau}) \label{34} \feqr \noi so
$Z_{A=0}(\tau , \bar{\tau}) d\tau d\bar{\tau}/\tau_2$ is modular
invariant. In order to avoid infinities coming from the momenta
integration an IR regularization procedure is needed. A possible way
to regularize the partition function without destroying the modular
invariance is to use an infrared cut-off $\epsilon$ and take the
limit $\epsilon \rightarrow 0$ after we divide by the singular term.
Doing this we get \beqr Z_{A=0}^R (\tau,\bar{\tau}) &=&
\lim_{\epsilon \rightarrow 0} \frac{\int_{\epsilon}^{M}dp_+
\int_{0}^{\infty} dp_- e^{-2\pi \tau_2 p_+ p_-}
\left[\hat{\eta}_{p_+ \tilde{S}_+}(\tau, \bar{\tau}) \hat{\eta}_{p_+
\tilde{S}_-}(\tau, \bar{\tau}) \right]^{-1}
}{\tau_2^2\int_{\epsilon}^{M}dp_+ \int_{0}^{\infty} dp_- e^{-2\pi
\tau_2 p_+ p_-} \left[\left(1-e^{-2\pi \tau_2 p_+ \tilde{S}_+}
\right)
\left(1-e^{-2\pi \tau_2 p_+ \tilde{S}_-} \right)\right]^{-1} } \non \\
&=& \frac{1}{\tau_2 \left( \sqrt{\tau_2} \eta(\tau) \bar{\eta
(\tau)}\right)^2}. \label{34.5} \feqr

\section{The partition function when $A \neq 0$}

When $A \neq 0$ the above calculation does
not lead to a modular invariant partition function. This is because
$A$ assumes the role of a worldsheet $O(2)$ gauge field in the $\sigma$-direction. 
It is possible to write the light-cone gauge fixed Hamiltonian
\rf{12} in the following form
\beqr
H &=& 2\pi p_+ p_- +\frac{1}{2\pi}\int d\sigma \Biggl[ \pi^2 (\mathcal{P}_i)^2 + (D_{\sigma}X_i)^2 \non \\
&-& p_+^2 \left( W_{1}(X^2_1 -X^2_2)-2W_2 X_1 X_2 -
\frac{1}{2}\Phi'' (X_1^2 +X_2^2)\right) \Biggl] \label{35} \feqr
\noi where $D^R_{\sigma}X_i = \partial_{\sigma}X_i -\frac{p_+ A}{6}
\epsilon_{il} X_l$. This suggests that $\frac{p_+ A}{6}
\epsilon_{ij}$ is an $O(2)$ connection along the $\sigma$-direction.
It is more convenient to define a complex spacetime coordinate to
turn the $O(2)$ connection to a $U(1)$ connection.  If $Z=X_1+iX_2$
then the Hamiltonian takes the form \beqr H = 2\pi p_+ p_- +
\frac{1}{2\pi}\int d\sigma \Biggl[ (2\pi)^2 P_z P_{\bar{z}} -
\frac{p_+^2}{2}W Z^2 -\frac{p_+^2}{2}\bar{W}\bar{Z}^2 +
\frac{p_+^2}{2}\Phi''Z\bar{Z} + D_{\sigma}Z
\overline{D_{\sigma}Z}\Biggr] \label{36} \feqr \noi where now
$D_{\sigma}=\partial_{\sigma}+i\frac{p_+ A}{6}$ and $W=W_1+iW_2$.
\par If we make the gauge transformation
\beqr Z=e^{-i\frac{p_+ A}{6}\sigma} \tilde{Z} \label{37} \feqr \noi
then the Hamiltonian becomes \beqr H = 2\pi p_+ p_- + \frac{1}{2\pi}
\int  d\sigma   \Biggl[ (2\pi)^2 P_{z} P_{\bar{z}}- \frac{p_+^2}{2}W
e^{-i\frac{p_+ A}{3}\sigma} \tilde{Z}^2 -\frac{p_+^2}{2}\bar{W}
e^{i\frac{p_+ A}{3}\sigma} \overline{\tilde{Z}}^2 +
\frac{p_+^2}{2}\Phi''\tilde{Z}\overline{\tilde{Z}} +
\partial_{\sigma} \tilde{Z}
\partial_{\sigma}\overline{\tilde{Z}}\Biggr].
\label{38} \feqr \noi The boundary condition satisfied by the gauge
transformed variable $\tilde{Z}$ is \beqr \tilde{Z} (\sigma + 2\pi,
t)= e^{i\frac{\pi}{3}p_+ A}\tilde{Z}(\sigma, t). \label{39} \feqr
\noi Expanding the fields \beqr
Z(\sigma, t) &=& \frac{1}{\sqrt{2}} \sum_{n \in Z} Z_n(t) e^{in \sigma} \non \\
P_{Z}(\sigma ,t) &=& \frac{1}{2\pi \sqrt{2}} \sum_{n \in Z}
P^{Z}_n(t) e^{-in\sigma} \label{40} \feqr \noi we get that \beqr
\tilde{Z}(\sigma, t) &=& \frac{1}{\sqrt{2}} \sum_{n \in Z} Z_n(t)
e^{i(n + \frac{p_+ A}{6})\sigma}. \label{40.5} \feqr \noi Now the
Hamiltonian becomes \beqr H  =2\pi p_+p_-+ \frac{1}{2} \sum_{n \in
Z} \Biggl[ P_n^ZP_n^{\bar{Z}}-\frac{p_+^2}{2}WZ_nZ_{-n}-
\frac{p_+^2}{2}\bar{W}\bar{Z}_n\bar{Z}_{-n}+\frac{p_+^2}{2}\Phi''Z_n\bar{Z}_n+(n+\frac{p_+A}{6})^2Z_n\bar{Z}_n\Biggr].
\label{41} \feqr \noi This is the same Hamiltonian as \rf{14} when
$A=0$ with the only difference that the term
$n^2((X_n^1)^2+(X_n^2)^2)$ is replaced by the term $(n+\frac{p_+
A}{6})^2((X_n^1)^2+(X_n^2)^2)$. Also note that the momentum operator
$\hat{\Pi}^i_n$ that appears in appendix A changes to \beqr
\hat{\Pi}_n^i =i(n+\frac{p_+A}{6}) \hat{X}_n^i \hat{P}_n^i -
i(n-\frac{p_+A}{6}) \hat{X}_{-n}^i \hat{P}_{-n}^i . \label{42} \feqr
This momentum operator generates $\sigma $ translations in
$\tilde{Z}$, $\overline{\tilde{Z}}$. This means that it generates
covariant $\sigma $ translations on $Z$, $\bar{Z} $. The only
difference is that $\pm n$ has been replaced by $(\pm
n+\frac{p_+A}{6})$. So the partition function becomes \beqr
Z_A(\tau, \bar{\tau })=C\int dp_+dp_-e^{-2\pi
\tau_2p_+p_-}\prod_{i=1}^2\prod_{n\in Z}
det^{-\frac{1}{2}}(D^i_{(n+\frac{p_+A}{6})}). \label{43} \feqr \noi
The product of determinants has been computed in appendix B so if we
define \beqr \tilde{\omega}^{(1,2)}_{n+ \frac{p_+
A}{6}}=\sqrt{(n+\frac{p_+ A}{6})^2 + p_+^2 S^2_{(+,-)}} \label{43.5}
\feqr \noi we have \beqr &&\prod_{n \in Z}det
(D^i_{(n+\frac{p_+A}{6})}) = \prod_{n \in Z} e^{2\pi
\tilde{\omega}^i_{(n+p_+A/6)} \tau_2}
(1-e^{-2\pi \tilde{\omega}^i_{(n+p_+A/6)} \tau_2 + 2i\pi (n+\frac{p_+A}{6}) \tau_1}) \non \\
&\cdot&(1-e^{-2\pi \tilde{\omega}^i_{(n+p_+A/6)} \tau_2 - 2i\pi
(n+\frac{p_+A}{6}) \tau_1}) = e^{4\pi \tau_2 \Delta_i(\frac{p_+
A}{6})} f_i[\frac{p_+ A}{6}] (\tau, \bar{\tau}) f_i[-\frac{p_+
A}{6}] (\tau, \bar{\tau}) \label{44} \feqr \noi where \beqr
\Delta^{(1,2)}_{\frac{p_+ A}{6}}&=& \Delta_{\frac{p_+ A}{6}}(p_+
S_{(+,-)}) =
\frac{1}{2}\sum_{n \in Z} \sqrt{(n+\frac{p_+ A}{6})^2 + p^2_+ \tilde{S}^2_{(+,-)}} \non \\
&=& -\frac{1}{2\pi^2}\sum_{n=1}^{\infty}\int_0^{\infty} ds e^{-n^2 s
-\frac{p^2_+ \tilde{S}^2_{(+,-)}\pi^2}{s}}
\cos (\pi \frac{p_+ A}{3}n) \non \\
f_i[\frac{p_+ A}{6}] (\tau, \bar{\tau}) &=& \prod_{n \in Z}
(1-e^{-2\pi \tilde{\omega}^i_{(n+p_+A/6)} \tau_2 + 2i\pi
(n+\frac{p_+A}{6}) \tau_1}). \label{45} \feqr \noi In terms of the
deformed theta functions of appendix D the partition function
becomes \beqr Z_A(\tau, \bar{\tau })=C\int dp_+dp_-e^{-2\pi
\tau_2p_+p_-}
\left( \frac{ \hat{\Theta}_{p_+ \tilde{S}_+}\left[ \begin{array}{c} \frac{1}{2}+\frac{p_+ A}{6} \\
\frac{1}{2} \end{array} \right] (\tau, \bar{\tau})
\hat{\Theta}_{p_+ \tilde{S}_-}
\left[ \begin{array}{c} \frac{1}{2}+\frac{p_+ A}{6} \\
\frac{1}{2} \end{array} \right] (\tau, \bar{\tau})
}{\hat{\eta}_{p_+ \tilde{S}_+}(\tau, \bar{\tau})
 \hat{\eta}_{p_+ \tilde{S}_-}(\tau, \bar{\tau}) }\right)^{-\frac{1}{2}}
\label{46} \feqr
\par This is modular invariant for fixed $A$ since the $p_-$ integration forces $p_+$ to be zero if we return to Minkowski space.
Furthermore it is equal to the free field partition function upon
regularization \cite{russotseytlin} \cite{hammou}.  Nevertheless, if
we sum over all possible twists before we carry out the $p_-$
integration, the partition function changes to \beqr Z(\tau,
\bar{\tau })=\int dC\frac{1}{\tau_2}
\left( \frac{ \hat{\Theta}_{R}\left[ \begin{array}{c} \frac{1}{2}+C \\
\frac{1}{2} \end{array} \right] (\tau, \bar{\tau})}
{\hat{\eta}_{R}(\tau, \bar{\tau})}\right)^{-1}=
 \int dC\frac{1}{\tau_2}
\frac{\left| \eta(\tau)\right|^{2} }
{ \left|\Theta\left[ \begin{array}{c} \frac{1}{2}+C \\
\frac{1}{2} \end{array} \right] (0,\tau )\right|^{2} }.
\label{46.5}
\feqr
This is not modular invariant since we
only have twists in the $\sigma$-direction. The
reason for this is the choice
of the light-cone gauge. Consider the action 
\rf{4} in the conformal gauge. This can be written in the form
\beqr 
S &=& \frac{1}{2\pi}\int
d\sigma dt \Biggl[ - \partial_{t} X^+
\partial_t X^- + \partial_{\sigma} X^+ \partial_{\sigma} X^-
+ (D_tX)_i (D_tX)_i - (D_{\sigma}X)_i (D_{\sigma}X)_i \non \\
&+& \tilde{F} \partial_t X^+ \partial_t X^+ - \tilde{F}
\partial_{\sigma} X^+ \partial_{\sigma} X^+ \Biggr] \label{47} \feqr
\noi where \beqr
(D_t X)_i &=& \partial_t X_i - \frac{1}{6} A_{ij}(X^+) \partial_{\sigma}X^+ X_j \non \\
(D_{\sigma} X)_i &=& \partial_{\sigma} X_i - \frac{1}{6} A_{ij}(X^+) \partial_{t}X^+ X_j \non \\
\tilde{F} &=& F + \frac{1}{36}A_{ij}(X^+)A_{il}(X^+) X_j X_l.
\label{48} \feqr
 \noi Note that \rf{3} implies that $\tilde{F}$
satisfies the equation \beqr
\partial^2_T \tilde{F}+ 2\Phi^{''}=0.
\label{49} \feqr \noi The action \rf{47} implies that we also have a
covariant derivative in the $t$-direction. In the light-cone gauge
$(D_t X)_i = \partial_t X_i$ so the symmetry in the $\sigma$- and
$t$-directions is broken. This explains why the partition function
we obtained is not modular invariant.  Upon the introduction of
twists in the $t$-direction, similar to the twists in the
$\sigma$-direction we get that
\beqr
Z(\tau, \bar{\tau })=
 \int_0^1 \int_0^1 dCdD\frac{1}{\tau_2}
\frac{\left| \eta(\tau)\right|^{2} }
{ \left|\Theta\left[ \begin{array}{c} \frac{1}{2}+C \\
\frac{1}{2}+D \end{array} \right] (0,\tau )\right|^{2} }.
\label{49.5}
\feqr
This is now explicitly modular invariant and it
is the partition function when we integrate over axion fields of the
form we have considered, as suggested by the covariant form of the
action.

\section{Conclusion}

In this paper motivated by the fascinating structure of potentially
exactly solvable plane wave backgrounds, we have investigated the
modular properties of the one-loop amplitude for the closed bosonic
string in four dimensions and in the presence of a metric, an axion
and a dilaton field. We found that when an axion field is switced
off ($A=0$) the partition function
is explicitly modular invariant.
In the case the axion field is switced on ($A\neq 0$) 
the partition function is not modular invariant since
we have a twist only in the $\sigma$-direction
which breaks the equivalence of the $\sigma$- and $t$-directions. 
One way to recover modular invariance is to
integrate over all the possible twists in both directions. However it seems
necessary to develop a covariant
quantization procedure to justify such an integration. Another way
to obtain the modular invariant partition function is through equivalences that
exist between the bosonic string in
certain pp-wave backgrounds and Wess-Zumino-Witten models based on
non-semisimple Lie groups of the type discussed in \cite{kiritsis}. 
\addcontentsline{toc}{subsection}{Appendix A }
\section*{Appendix A }
\renewcommand{\theequation}{A.\arabic{equation}}
\setcounter{equation}{0}

Let us consider the generators of $t$ and $\sigma$ translations on
the n-mode part of the field operator $\hat{X}^i(t,\sigma)$
expansion, $\hat{X}^i_n (t) e^{in\sigma} + \hat{X}^i_{-n} (t)
e^{-in\sigma}$ for $n>0$. In this appendix to simplify notation we
omit the tilde from the $\hat{X}^i_n$ and the $\hat{P}^i_n$. The
generator of $t$ translations is just the harmonic oscillator
Hamiltonian $\hat{H}^i_n$ while the generator of $\sigma$
translations is \beqr \hat{\Pi}^i_n=in \left( \hat{X}^i_n
\hat{P}^i_{n} - \hat{X}^i_{-n} \hat{P}^i_{-n} \right). \label{a1}
\feqr \noi It satisfies \beqr [\hat{\Pi}^i_n,\hat{X}^j_m (t)
e^{im\sigma} + \hat{X}^j_{-m} (t)
e^{-im\sigma}]=-i\delta^{ij}\delta_{nm}\frac{\partial}{\partial
\sigma} \left(\hat{X}^i_n (t) e^{in\sigma} + \hat{X}^i_{-n} (t)
e^{-in\sigma} \right). \label{a2} \feqr \noi We want to compute the
partition function of the oscillators on the torus. To do this we
propagate oscillator states from $\sigma$, $t-\pi \tau_2$ to $\sigma
+ 2\pi \tau_1$, $t+\pi\tau_2$ along a path in the worldsheet which
we discretize. Now we have \beqr && <X^{i,j+1}_{\pm n},
\sigma_{j+1}, t_{j+1} | X^{i,j}_{\pm n}, \sigma_{j}, t_{j} > =
<X^{i,j+1}_{\pm n}|e^{-i\hat{H}^i_n \delta t} e^{i\hat{\Pi}^i_n \delta \sigma} |X^{i,j}_{\pm n} > \non \\
&=& \delta(X^{i,j+1}_n - X^{i,j}_n) \delta(X^{i,j+1}_{-n} -
X^{i,j}_{-n}) - i \delta t
<X^{i,j+1}_{\pm n}|\hat{H}^i_n|X^{i,j}_{\pm n}> \non \\
&+& i \delta \sigma <X^{i,j+1}_{\pm n}|\hat{\Pi}^i_n|X^{i,j}_{\pm
n}>+ quad. \label{a3} \feqr \noi Now observe that \beqr
\delta(X^{i,j+1}_n - X^{i,j}_n) \delta(X^{i,j+1}_{-n} -
X^{i,j}_{-n})= \frac{1}{(2\pi)^2} \int dP^{i,j}_n dP^{i,j}_{-n}
e^{iP^{i,j}_{n} (X^{i,j+1}_{n} - X^{i,j}_{n})} e^{iP^{i,j}_{-n}
(X^{i,j+1}_{-n} - X^{i,j}_{-n})}, \label{a4} \feqr \beqr <X^{i,
j+1}_{\pm n}|P^{i,j}_{\pm n}>=\frac{1}{2\pi}e^{iP^{i,j}_n
X^{i,j+1}_{n}}e^{iP^{i,j}_{-n} X^{i,j+1}_{-n}} \label{a4.5} \feqr
\beqr <X^{i,j+1}_{\pm n}|\hat{H}^i_n|X^{i,j}_{\pm n}> &=& \int
dP^{i,j}_n dP^{i,j'}_n dP^{i,j}_{-n} dP^{i,j'}_{-n} <X^{i,j+1}_{\pm
n}|P^{i,j'}_{\pm n}> <P^{i,j'}_{\pm n}|\hat{P}^i_n
\hat{P}^i_{-n}|P^{i,j}_{\pm n}>
<P^{i,j}_{\pm n}|X^{i,j}_{\pm n}> \non \\
&+& (\omega^i_n)^2 <X^{i,j+1}_{\pm n}|\hat{X}^i_n \hat{X}^i_{-n}|X^{i,j}_{\pm n}> \non \\
&=& \frac{1}{(2\pi)^2} \int dP^{i,j}_n dP^{i,j}_{-n} e^{iP^{i,j}_{n}
(X^{i,j+1}_{n} - X^{i,j}_{n})} e^{iP^{i,j}_{-n} (X^{i,j+1}_{-n} -
X^{i,j}_{-n})} H^i_n (P^{i,j}_{\pm n}, X^{i,j}_{\pm n}) \label{a5}
\feqr \noi and similarly \beqr
&& <X^{i,j+1}_{\pm n}|\hat{\Pi}^i_n|X^{i,j}_{\pm n}> = in \int dP^{i,j}_n dP^{i,j'}_n dP^{i,j}_{-n} dP^{i,j'}_{-n}\non \\
&& \Biggl[ X^{i,j}_{n} <X^{i,j+1}_{\pm n}|P^{i,j'}_{\pm n}>
<P^{i,j'}_{\pm n}|\hat{P}^i_{n}|P^{i,j}_{\pm n}>
<P^{i,j}_{\pm n}|X^{i,j}_{\pm n}> \non \\
&-& X^{i,j}_{-n} <X^{i,j+1}_{\pm n}|P^{i,j'}_{\pm n}> <P^{i,j'}_{\pm
n}|\hat{P}^i_{-n}|P^{i,j}_{\pm n}>
<P^{i,j}_{\pm n}|X^{i,j}_{\pm n}> \Biggl] \non \\
&=&  \frac{1}{(2\pi)^2} \int dP^{i,j}_n dP^{i,j}_{-n}
e^{iP^{i,j}_{n} (X^{i,j+1}_{n} - X^{i,j}_{n})} e^{iP^{i,j}_{-n}
(X^{i,j+1}_{-n} - X^{i,j}_{-n})} \Pi^i_n (P^{i,j}_{\pm n},
X^{i,j}_{\pm n}) \label{a6} \feqr \noi Putting together \rf{a4},
\rf{a5} and \rf{a6} we get \beqr
&&<X^{i,j+1}_{\pm n}|e^{-i\hat{H}^i_n \delta t} e^{i\hat{\Pi}^i_n \delta \sigma} |X^{i,j}_{\pm n} > \non \\
&=& \frac{1}{(2\pi)^2} \int dP^{i,j}_n dP^{i,j}_{-n} e^{iP^{i,j}_{n}
(X^{i,j+1}_{n} - X^{i,j}_{n})} e^{iP^{i,j}_{-n} (X^{i,j+1}_{-n} -
X^{i,j}_{-n})} \left( 1- i\delta t H^i_n (P^{i,j}_{\pm n},
X^{i,j}_{\pm n}) + i\delta \sigma
\Pi^i_n (P^{i,j}_{\pm n}, X^{i,j}_{\pm n})\right) \non \\
&=& \frac{1}{(2\pi)^2} \int dP^{i,j}_n dP^{i,j}_{-n} e^{iP^{i,j}_{n}
(X^{i,j+1}_{n} - X^{i,j}_{n})} e^{iP^{i,j}_{-n} (X^{i,j+1}_{-n} -
X^{i,j}_{-n})} e^{- i\delta t H^i_n (P^{i,j}_{\pm n}, X^{i,j}_{\pm
n}) }e^{i\delta \sigma \Pi^i_n (P^{i,j}_{\pm n}, X^{i,j}_{\pm n})} +
quad. \label{a7} \feqr \noi Now, since the momentum and the
Hamiltonian commute, the partition function is independent of the
path by which we join ($\sigma$, $t-\pi\tau_2$) to ($\sigma +
2\pi\tau_1$, $t+\pi\tau_2$) so we can take this path to be a
straight line. Hence we have that $\delta \sigma =
\frac{\tau_1}{\tau_2} \delta t$ and then \beqr
Z^i_n &\equiv& <X^{i,final}_n, \sigma + 2\pi \tau_1, t+\pi \tau_2|X^{i,initial}_n, \sigma, t -\pi \tau_2> \non \\
&=& C \int \prod_j dX^{i,j}_{n} dX^{i,j}_{-n}dP^{i,j}_{n}
dP^{i,j}_{-n}
e^{iP^{i,j}_{n} (X^{i,j+1}_{n} - X^{i,j}_{n})} e^{iP^{i,j}_{-n} (X^{i,j+1}_{-n} - X^{i,j}_{-n})} \non \\
&& e^{- i\delta t H^i_n (P^{i,j}_{\pm n}, X^{i,j}_{\pm n}) }e^{i
\frac{\tau_1}{\tau_2}\delta t \Pi^i_n (P^{i,j}_{\pm n}, X^{i,j}_{\pm
n})}. \label{a8} \feqr \noi In the continuum limit this becomes
\beqr Z^i_n = C \int DX^{i}_{n} DX^{i}_{-n}DP^{i}_{n} DP^{i}_{-n}
e^{i\int \left(P^i_{n}\dot{X}^i_n + P^i_{-n}\dot{X}^i_{-n} -H^i_n
(P^{i}_{\pm n}, X^{i}_{\pm n})+ \frac{\tau_1}{\tau_2} \Pi^i_n
(P^{i}_{\pm n}, X^{i}_{\pm n}) \right)dt}. \label{a9} \feqr \noi
Analytically continuing to imaginary time by letting $t\rightarrow
-it$, $\tau_2 \rightarrow -i\tau_2$ and performing the $P^i_{\pm n}$
integrations we get \beqr Z^i_n &=& C \int DX^i_n DX^i_{-n} e^{-\int
X^i_n \left(-\frac{d^2}{dt^2}+2in\frac{\tau_1}{\tau_2}\frac{d}{dt} +
n^2
\frac{\tau_1^2}{\tau_2^2} + (\omega^i_n)^2 \right) X^i_{-n} dt } \non \\
&=& C
det^{-1}\left(-\frac{d^2}{dt^2}+2in\frac{\tau_1}{\tau_2}\frac{d}{dt}
+ n^2 \frac{\tau_1^2}{\tau_2^2} + (\omega^i_n)^2 \right).
\label{a10} \feqr
\par For the zero mode $X^i_0$ formula \rf{a10} is no longer valid because we do not have a complex pair of modes,
but instead we have a real mode. Doing a similar calculation we get
\beqr Z^i_0 &=& C \int DX^i_0 e^{-\frac{1}{2}\int X^i_0
\left(-\frac{d^2}{dt^2}+ (\omega^i_0)^2 \right) X^i_{0} dt } = C
det^{-\frac{1}{2}}\left(-\frac{d^2}{dt^2}+ (\omega^i_0)^2 \right).
\label{a11} \feqr


\addcontentsline{toc}{subsection}{Appendix B }
\section*{Appendix B }
\renewcommand{\theequation}{B.\arabic{equation}}
\setcounter{equation}{0}

Consider now the operator \beqr D_{(n+a)} =
-\frac{d^2}{dt^2}+2i(n+a)\frac{\tau_1}{\tau_2}\frac{d}{dt} + (n+a)^2
\frac{\tau_1^2}{\tau_2^2} + \tilde{\omega}^2_{(n+a)} \label{b1}
\feqr \noi acting on periodic functions on $(-\pi \tau_2, \pi
\tau_2)$. The basis of eigenfuctions is \beqr
f_m(t)=e^{im\frac{t}{\tau_2}}, \quad m\in Z. \label{b2} \feqr \noi
The corresponding eigenvalues are \beqr \lambda_{m}^{n+\alpha} =
\frac{\left(m-(n+a)\tau_1
\right)^2}{\tau^2_2}+\tilde{\omega}^2_{(n+a)}. \label{b3} \feqr \noi
In computing the determinant of $D_n$ we are going to use the zeta
function regularization. This means that \beqr
\prod_{n=-\infty}^{\infty} \alpha =\alpha^{2\zeta (0)+1}=1, \qquad
\prod_{n=1}^{\infty} n^{\alpha}= e^{-\alpha \zeta'(0)}=
(2\pi)^{\frac{\alpha}{2}} \label{b4} \feqr \noi and \beqr
\prod_{n=-\infty}^{\infty}(n+\alpha)=\alpha \prod_{n=1}^{\infty}
(-n^2)\left(1-\frac{\alpha^2}{n^2} \right) =2i\sin \pi \alpha.
\label{b5} \feqr \noi Now \beqr det(D_{(n+a)}) &=& \prod_{m \in Z}
\lambda^{n+\alpha}_m = \prod_{m \in Z} \frac{\tau^2_2
\tilde{\omega}^2_{(n+a)}
+ (m-(n+a)\tau_1)^2}{\tau^2_2} \non \\
&=& \prod_{m \in Z} (m-(n+a)\tau_1 -i\tilde{\omega}_{(n+a)}\tau_2 )(m-(n+a)\tau_1 +i\tilde{\omega}_{(n+a)}\tau_2 ) \non \\
&=& -4\sin [\pi((n+a)\tau_1 + i\tilde{\omega}_{(n+a)} \tau_2)]
\sin [\pi((n+a)\tau_1 - i\tilde{\omega}_{(n+a)} \tau_2)] \non \\
&=&-e^{2\pi \tilde{\omega}_{(n+a)}\tau_2}\left( 1-e^{-2\pi
\tilde{\omega}_{(n+a)}\tau_2+2i\pi(n+a)\tau_1}\right) \left(
1-e^{-2\pi \tilde{\omega}_{(n+a)}\tau_2- 2i\pi (n+a)\tau_1}\right).
\label{b6} \feqr

\addcontentsline{toc}{subsection}{Appendix C }
\section*{Appendix C }
\renewcommand{\theequation}{C.\arabic{equation}}
\setcounter{equation}{0} Consider \beqr F(a)=\sum_{n\in Z}((n+a)^2 +
c^2)^p. \label{d2} \feqr It is possible to write $F(a)$ as \beqr
F(a)=\int_{-\infty}^{\infty}\delta(y-a)F(y)dy = \sum_{k\in
Z}e^{2i\pi ka}\int_{-\infty}^{\infty}e^{-2i\pi k y}(y^2 + c^2)^p dy.
\label{d3} \feqr \noi Substituting the expression \beqr (y^2
+c^2)^p=\frac{1}{\Gamma(-p)}\int_{0}^{\infty}t^{-(1+p)}e^{-(y^2
+c^2)t}dt \label{d4} \feqr \noi into \rf{d3} and performing the
Gaussian integration we get \beqr F(a)&=&
\frac{\sqrt{\pi}}{\Gamma(-p)}\sum_{k\in Z}e^{2i\pi
ka}\int_{0}^{\infty}t^{-(\frac{3}{2}+p)}e^{-(tc^2 + \frac{\pi^2
k^2}{t})}dt
\non \\
&=&
\frac{\sqrt{\pi}}{\Gamma(-p)}\Gamma(-p-\frac{1}{2})c^{1+2p}+\frac{\sqrt{\pi}}{\Gamma(-p)}\sum_{k=1}^{\infty}2\cos
(2k\pi a) \int_{0}^{\infty} t^{-(\frac{3}{2}+p)} e^{-(tc^2 +
\frac{\pi^2 k^2}{t})}dt. \label{d5} \feqr \noi Substituting in
formula \rf{d2} and expressing the result in terms of the modified
Bessel function we have \beqr F(a)= \frac{\sqrt{\pi}}{\Gamma(-p)}
c^{1+2p}\left( \Gamma(-p-\frac{1}{2})+ \sum_{k=1}^{\infty}4(\pi k
c)^{-p-\frac{1}{2}}\cos (2k\pi c)K_{-p-\frac{1}{2}}(2k\pi c)\right).
\label{d6} \feqr \noi In the particular case $a=0$ and using the
symmetry $n \rightarrow -n$ we get \beqr \sum_{n=1}^{\infty}(n^2
+c^2)^{p}&=& \frac{1}{2} \left[\sum_{n \in Z}(n^2 + c^2)^p -
c^{2p}\right]
=\frac{1}{2}\left[F(0)-c^{2p}\right] \non \\
&=& -\frac{c^{2p}}{2}+ \frac{\sqrt{\pi}}{2\Gamma(-p)} c^{1+2p}\left(
\Gamma(-p-\frac{1}{2})+ \sum_{k=1}^{\infty}4(\pi k
c)^{-p-\frac{1}{2}}K_{-p-\frac{1}{2}}(2k\pi c)\right). \label{d1}
\feqr
\addcontentsline{toc}{subsection}{Appendix D }
\section*{Appendix D }
\renewcommand{\theequation}{D.\arabic{equation}}
\setcounter{equation}{0}

It is possible to define a deformed version of the product $\eta
(\tau) \overline{\eta (\tau)}$ where $\eta (\tau)$ is the Jacobi
$\eta$-function in such a way that it transforms simply under
modular transformations. Define \beqr \hat{\eta}_c(\tau,
\bar{\tau})= e^{2\pi \tau_2 \Delta(c)}\prod_{n \in Z} (1- e^{-2\pi
\tau_2 \sqrt{n^2 + c^2} + 2i\pi n \tau_1}) \label{c1} \feqr \noi
where \beqr
\Delta(c)=-\frac{1}{2\pi^2}\sum_{n=1}^{\infty}\int_{0}^{\infty}ds
e^{-n^2 s - \frac{c^2 \pi^2}{s}}. \label{c2} \feqr \noi The modular
properties satisfied by $\hat{\eta}_{c}(\tau , \bar{\tau}) $ are
given by \beqr \label{c2.5}
\hat{\eta}_{c}(\tau +1, \bar{\tau}+1) &=&  \hat{\eta}_{c}(\tau, \bar{\tau})  \non \\
\hat{\eta}_{c}(-\frac{1}{\tau}, -\frac{1}{\bar{\tau}}) &=&
\hat{\eta}_{c/|\tau|}(\tau, \bar{\tau}). \feqr \noi Note that in
particular $\Delta(0)=-1/12$. When $c\rightarrow 0$,
$\hat{\eta}_c(\tau, \bar{\tau})\rightarrow 0$, but it is possible to
regularize it by dropping the factor that goes to zero. In this way
we obtain \beqr \hat{\eta}^R_0 (\tau, \bar{\tau})\equiv
\lim_{c\rightarrow 0} \frac{\hat{\eta}_c(\tau,
\bar{\tau})}{(1-e^{-2\pi c \tau_2})}= \eta(\tau) \overline{\eta
(\tau)}. \label{c3} \feqr \noi It is also possible to define
deformed $\theta$-function bilinears. Define \beqr \hat{\Theta}_c
\left[ \begin{array}{c} a \\ b \end{array} \right] (\tau,
\bar{\tau})&=& e^{2\pi\tau_2 \hat{\Delta}_{a+\frac{1}{2}}(c)}
\prod_{n \in Z}(1-e^{-2\pi \tau_2 \sqrt{n^2 +c^2} + 2i\pi n \tau_1}) \non \\
&& (1-e^{-2\pi \tau_2 \sqrt{(n+a+1/2)^2 +c^2} + 2i\pi (n+a+1/2) \tau_1 + 2i\pi (b+\frac{1}{2})}) \non \\
&& (1-e^{-2\pi \tau_2 \sqrt{(n+a+1/2)^2 +c^2} - 2i\pi (n+a+1/2)
\tau_1 - 2i\pi (b+\frac{1}{2})}) \label{c4} \feqr \noi where \beqr
\hat{\Delta}_a (c)= -\frac{1}{2\pi^2}
\sum_{n=1}^{\infty}\int_{0}^{\infty} ds e^{-n^2 s - \frac{c^2
\pi^2}{s}}(1+2\cos(2\pi a n)). \label{c5} \feqr \noi The modular
transformations of the deformed $\Theta$-functions are given by
\beqr \hat{\Theta }_c \left[ \begin{array}{c} a \\ b \end{array}
\right] (-\frac{1}{\tau}, -\frac{1}{\bar{\tau}})&=&
\hat{\Theta }_{\frac{c}{|\tau|}} \left[ \begin{array}{c} b \\ -a \end{array} \right] (\tau, \bar{\tau}) \non \\
\hat{\Theta }_c \left[ \begin{array}{c} a \\ b \end{array} \right]
(\tau +1, \bar{\tau} +1)&=& \hat{\Theta }_{c} \left[
\begin{array}{c} a \\ a+b+\frac{1}{2} \end{array} \right] (\tau,
\bar{\tau}). \label{c6} \feqr \noi Again, when $c\rightarrow 0$,
$\hat{\Theta}_c \left[ \begin{array}{c} a \\ b \end{array} \right]
(\tau, \bar{\tau})\rightarrow 0$, but it is possible to regularize
it by dropping the factor that goes to zero. In this way we obtain
\beqr \hat{\Theta}_0^R  \left[ \begin{array}{c} a \\ b \end{array}
\right] (\tau, \bar{\tau}) = \lim_{c\rightarrow 0}
\frac{\hat{\Theta}_c  \left[ \begin{array}{c} a \\ b \end{array}
\right] (\tau, \bar{\tau}) } {1-e^{-2\pi c \tau_2}}. \label{c7}
\feqr \noi One can easily prove that this regularized deformed
$\Theta$-function is given by \beqr \hat{\Theta}_0^R  \left[
\begin{array}{c} a \\ b \end{array} \right] (\tau, \bar{\tau}) =
\Theta  \left[ \begin{array}{c} a \\ b \end{array} \right] (0, \tau)
\overline{\Theta  \left[ \begin{array}{c} a \\ b \end{array} \right]
(0, \tau)} \label{c8} \feqr \noi where the $\Theta$-functions that
appear on the righthand side of \rf{c8} are the usual theta
functions with characteristics. It is worth mentioning that in this
context the factor $\sqrt{-i\tau}$ that appears in the modular
transformation properties of the usual $\Theta$-functions can be
attributed to the necessary regularization when $c\rightarrow 0$
since \beqr \hat{\Theta}_0^R  \left[ \begin{array}{c} a \\ b
\end{array} \right] (-\frac{1}{\tau}, -\frac{1}{\bar{\tau}}) =
\lim_{c\rightarrow 0}
\frac{1-e^{-2\pi\frac{c}{|\tau|}\tau_2}}{1-e^{-2\pi c
\frac{\tau_2}{|\tau|^2} }} \hat{\Theta}_0^R  \left[ \begin{array}{c}
b \\ -a \end{array} \right] (\tau, \bar{\tau}) =
|\tau|\hat{\Theta}_0^R  \left[ \begin{array}{c} b \\ -a \end{array}
\right] (\tau, \bar{\tau}). \label{c9} \feqr \noi Furthermore the
deformed $\Theta$-functions have the following symmetry properties
\beqr \hat{\Theta}_c\left[ \begin{array}{c} a+1 \\ b \end{array}
\right] (\tau, \bar{\tau})= \hat{\Theta}_c\left[ \begin{array}{c} a
\\ b+1 \end{array} \right] (\tau, \bar{\tau})= \hat{\Theta}_c\left[
\begin{array}{c} a \\ b \end{array} \right] (\tau, \bar{\tau})
\label{c10} \feqr \noi and \beqr \hat{\Theta}_c\left[
\begin{array}{c} -a \\ -b \end{array} \right] (\tau, \bar{\tau})=
\hat{\Theta}_c\left[ \begin{array}{c} a \\ b \end{array} \right]
(\tau, \bar{\tau}). \label{c11} \feqr
\bibliographystyle{plain}

\end{document}